\documentstyle[preprint,aps,epsfig]{revtex}
\draft
\begin{document}
\preprint{}
\title{\bf Differential Transverse Flow in Central C-Ne and C-Cu Collisions at 3.7 GeV/nucleon}
\bigskip
\author{\bf L. Chkhaidze, T. Djobava\footnote{email: djobava@sun20.hepi.edu.ge},
 L. Kharkhelauri}
\address{High Energy Physics Institute, Tbilisi State University\\
University St 9, 380086 Tbilisi, Republic of Georgia}
\bigskip\bigskip
\author{\bf Bao-An Li\footnote{email: Bali@astate.edu}}
\address{Department of Chemistry and Physics\\
P.O. Box 419, Arkansas State University\\
State University, Arkansas 72467-0419, USA}
\maketitle

\begin{quote}
Differential transverse flow of protons and pions in central C-Ne and C-Cu
collisions at a beam energy of 3.7 GeV/nucleon was measured as a function
of transverse momentum
at the SKM-200-GIBS setup of JINR. In agreement with predictions of a transversely moving
thermal model, the strength of proton differential transverse flow
is found to first increase gradually and then saturate with the increasing
transverse momentum in both systems. While pions are preferentially emitted in the same
direction of the proton transverse flow in the reaction of C-Ne, they exhibit
an anti-flow to the opposite direction of the proton transverse flow in the
reaction of C-Cu due to stronger shadowing effects of the heavier target in the whole
range of transverse momentum.
\end{quote}

\newpage

\section{INTRODUCTION }

The ultimate goal of high energy heavy-ion studies is to investigate
nuclear matter under extreme conditions of high density and temperature.
In particular, extensive experimental and theoretical efforts have been
devoted to probe the nuclear Equation of State (EOS), to identify the
Quantum Chromodynamics (QCD) phase transition and to study properties of the
Guark-Gluon Plasma (QGP) which is a novel form of matter. New information
extracted from these studies is critical to many interesting
questions in both nuclear physics and astrophysics. It is thus
fundamentally important to find sensitive experimental probes of hot and
dense nuclear matter. The collective phenomena found in heavy-ion collisions,
such as the transverse (sideward directed) and elliptic flow of nuclear matter,
have been shown to be among the most sensitive probes of the nuclear EOS.
The transverse flow is the sideward deflection of nuclear matter moving
forward and backward in the reaction plane. The transverse flow
has been observed for nucleons, nuclear fragments and newly produced
particles, such as pions and kaons, in heavy-ion collisions at all
available energies ranging from the Fermi energy to the highest
RHIC energy\cite{gary,hhg,reis,ogi,rai,rhic}. Especially,
around 4 GeV/nucleon where a transition from ``squeeze-out" perpendicular
to the reaction plane to the in-plane flow is expected, a number of flow
measurements have been carried out at the AGS/BNL\cite{eos,e917} and
the JINR/DUBNA\cite{dubna1,dubna2,dubna3,dubna4,dubna5}.
Clear evidence of the transverse and elliptic flow effects for protons
and $\pi^{-}$ mesons have been obtained in central C-Ne and C-Cu
collisions at energy of 3.7 GeV/nucleon by SKM-200-GIBS Collaboration at
JINR/DUBNA. In C-Ne interactions the transverse flow of $\pi^{-}$ mesons
is in the same direction as for the protons, while in C-Cu collisions pions
show antiflow behabiour. From the transverse momentum and azimuthal
distributions of protons and $\pi^{-}$ mesons with respect to the reaction
plane defined by participant protons, the flow $F$ (a measure of the
collective transverse momentum transfer in the reaction plane) and the parameter
$a_{2}$ (a measure of the anisotropic emission strength) have been extracted.
The flow effects increase with the mass of the particle and the mass number
of target $A_{T}$. The scaled flow $F_{s}=F/(A_{P}^{1/3}+A_{T}^{1/3})$ have been used
for comparison of transverse flow results of SKM-200-GIBS with flow data
for various energies and projectile/target configurations.
The $F_{s}$ demonstrates a common scaling behaviour of protons
flow values for different energies (Bevalac, GSI/SIS, Dubna, AGS, SPS)
and systems. Focusing on the total transverse flow, i.e., integrated over transverse 
momentum, these studies have revealed much interesting physics. In the present work, we
report results of a differential transverse flow analysis as a function of transverse
momentum for protons and pions in central C-Ne and C-Cu collisions at a beam energy of
3.7 Gev/nucleon measured at the SKM-200-GIBS setup of JINR.

\section{EXPERIMENT}

The SKM-200-GIBS setup consists of a 2 m streamer chamber with
volume $2 \times 1 \times 0.6~ m^{3}$, placed in a magnetic field
of 0.8 Tesla, and a triggering system. The streamer chamber was exposed to a
beam of C nuclei accelerated in the synchrophasotron
up to a momentum of 4.5 GeV/c/nucleon (beam energy $E_{beam}= 3.7$ GeV/nucleon).
The thickness of the solid target Cu (in the form of a thin disc) was
0.2 g/cm$^{2}$. Neon gas filling of the chamber also served as a nuclear target.
Photographs of the events were taken using an optical system with three
objectives. The experimental set-up and the logic of the triggering system
are presented in Fig. 1. 

The triggering system allowed the selection of
``inelastic" and ``central" collisions. The ``inelastic" trigger, consisting of two 
sets of scintillation counters mounted upstream (S$_{1}$ - S$_{4}$) and 
downstream (S$_{5}$, S$_{6}$) the chamber,
has been selecting all inelastic interactions of incident nuclei on a target.
The ``central" triggering system was consisting of the same upstream part as in the
``inelastic" system and of scintillation veto counters (S$_{ch}$, S$_{n}$),
registering a projectile and its charged and neutral spectator fragments,
in the downstream part. All counters were made from the plastic scintillators and
worked with photomultipliers PM-30. The S$_{1}$ counter with the scintillator
of 20$\times$20$\times$0.5 cm$^{3}$ size, worked in the amplitude regime and
identified the beam nuclei by the charge. The nuclei from the beam, going
to the target, has been selected using the profile counters S$_{2}$, S$_{3}$
with the plastic of 15 mm diameter and 3 mm thickness and ``thin" counter
S$_{4}$ (15 mm and 0.1 mm correspondly). The S$_{ch}$ counters (two
counters with plastic of 25$\times$25$\times$0.5 cm$^{3}$ size) were placed
at a distance of 6 m downstream the target
 and registered the secondary charged particles, emitted from the target
within a cone of half angle $\Theta_{ch}$=$2.4^{0}$, $2.9^{0}$
(the trigger efficiency was 99$\%$ for events with a single charged particle in the cone).
The  S$_{n}$ counters registered the neutrons, emitted from the target in the solid angle
$\Theta_{n}$=$1.8^{0}$, $2.8^{0}$
(the trigger efficiency was 80$\%$ for one neutron).
The S$_{n}$ telescope consists from the five counters
of 40$\times$40$\times$2 cm$^{3}$ size, layered by iron blocks of 10 cm
thickness.  The trigger selected central C-Ne and  C-Cu collisions defined
as those without charged projectile spectator fragments (with  $p_Z>3$ GeV/c ) 
within a cone of  half angle $\Theta_{ch}$ = 2.4$^{0}$.

The streamer chamber pictures were scanned twice, and a subsequent third
scan resolved ambiguities or discrepancies between two scans.
Primary results of scanning and measurements were biased due to several
experimental effects and appropriate corrections were introduced
\cite{exp1,exp2}:\\
1) The triggering system selected required projectile nuclei from
a primary beam with an efficiency higher than $99\%$. The contamination
due to interactions of other projectile nuclei with charge less than
required biased the values of average multiplicities of secondary
particles only insignificantly  (the correction value $\leq 0.5 \%$).\\
2) A trigger bias for central collisions arises whenever a projectile
fragment hits the minimum bias veto-counter system and thus simulates a
projectile-nucleus fragment. The effect was studied by simulating trajectories
of secondary particles generated within the framework of the cascade
model \cite{gud}. The geometry of the experimental set-up and magnetic field
distribution were taken into account. The biases thus estimated turned
out to be below statistical errors in multiplicity,
 $p_{t}$ and $y$ distributions.\\
3) The corrections due to secondary interactions within a solid target
turned out to be negligible (the correction value $\leq 0.5 \%$).\\
4) Pion detection was biased against low momenta. Pions were registered
with practically no bias when their momenta were $p \geq$ 40 MeV/c for the
solid targets and $p \geq$ 20 MeV/c for gaseous $Ne$. Pions with lower momenta
were also registered, however their detection efficiency depends on the path
length in the target. The corrections could be delivered properly only
for colliding with equal masses under the requirement of the symmetry
$y - p_{t}$ plots in the c.m. system ( $\sim 0.6 \%$ in average multiplicity).
A very rough estimation for asymmetric pairs of colliding nuclei showed that
the correction values are $(0.5- 2)\%$.\\
5) The samples of negative secondaries include $K^{-}$ and $\Sigma^{-}$
tracks.  The contamination was estimated from $pp$ data \cite{ben} and
from our results on strange particle production \cite{anik}
(the correction value $\leq 0.5 \%$).\\
6) Corrections for scanning losses originate from two sources:\\
a) scanning inefficiency ($\sim 2\%$);\\
b) losses of the tracks with a small projection length (which may be screened
by the target container or a flash around the vertex) and/or a small track
curvature ( $\sim 1-4 \%$) The corrections were estimated under the
requirement of azimuthal symmetry for each emission angle interval for all
groups of the measured interactions separately. The correction values
are $0.5-3.5 \%$.\\
7) The measurement precision essentially depends on the track length and
its dip angle. In the samples of events considered $\sim 5- 15\%$
of $\pi^{-}$ mesons turned out to be unmeasurable or were rejected because
of large errors. The measurement losses of $\pi^{-}$ mesons concerned
practically only well determined intervals of emission angles, and,
consequently, appropriate corrections, based on azimuthal symmetry, could
be introduced in the spectra. Inaccuracy of the corrections was calculated,
and systematic uncertainties of these corrections were found to be less than
statistical ones. The ratio $\sigma_{cent}$/$\sigma_{inel}$ (that characterizes the
centrality of selected events) is (9$\pm$1)$\%$ for C-Ne and (21$\pm$3)$\%$
for C-Cu. The average errors in measuring the momentum and production angle
are $<\Delta p/p>$= (8-10)$\%$ and $\Delta$$\Theta$ =1$^{0}-2^{0}$, respectively,
for protons, while for pions $<\Delta p/p>$= 5$\%$ and
$\Delta$$\Theta$ =0.5$^{0}$, respectively.

\section{DIFFERENTIAL FLOW  OF
NUCLEONS AND PIONS IN C-Ne AND C-Cu COLLISIONS }

Several different methods have been utilized for analyzing flow effects
in relativistic nuclear collisions, among them the transverse momentum
analysis \cite{pawel} and Fourier expansion of particle azimuthal angle
distributions \cite{ell1,ell2} have been most widely used. More recently,
it has been found that the differential analysis of the flow strength as a function
of transverse momentum is more useful in revealing detailed properties of the
hot and dense matter, see, e.g. \cite{rhic,bali1,bali2,gyu,fopi}.
In this work, we adopt the method of differential flow analysis
first used by the E877 collaboration in ref. \cite{e877}.
Let $N^{+}(N^{-})$ be the number of particles
emitted in the same (opposite) direction of the transverse flow near projectile
rapidity, then the ratio
$R(p_{t})=(dN^{+}/dp_{t})/(dN^{-}/dp_{t})$ as a
function of $p_{t}$  is a direct measure of the strength of differential flow
near the projectile rapidity. It has been shown experimentally that more detailed
information about the collective flow can be obtained by studying this ratio \cite{e877}.
The differential flow data also provides a much more stringent test ground for relativistic
heavy-ion reaction theories. It was shown theoretically by Li et al in ref. \cite{bali3}
and Voloshin in ref. \cite{vio} that the ratio $R(p_t)$ at high $p_{t}$
is particularly useful in studying the EOS of dense and hot matter formed in relativistic
heavy-ion collisions.

The data have been analysed event by event  using
the transverse momentum technique of P.Danielewicz and G.Odyniec
\cite{pawel}. The reaction plane have been defined
for the paricipant protons i.e. protons which are not fragments
of the projectile ($p/z >3$ GeV/c , $\Theta < 4^{0}$) and target
($p/z <0.2$ GeV/c). They represent the protons participating in the collision.
The analysis have been carried out on 723 C-Ne and 663 events. The average
multiplicities of participant protons are $<N_{p}> = 12.4 \pm 0.5$ in C-Ne and 
$<N_{p}> = 19.5 \pm 0.6$ in C-Cu, respectively.

The admixture of $\pi^{+}$ mesons
amongst the protons is about (25$\div$27) $\%$.
For the event by event analysis it is necessary to perform an
identification of $\pi^{+}$ mesons and separate them from the protons.
The statistical method have been used for identification of $\pi^{+}$ mesons.
The main assumption is based on the similarity of spectra of
$\pi^{-}$ and $\pi^{+}$ mesons ($n_{\pi}$, $p_{t}$, $p_{l}$). The
two-dimentional -- transverse and longitudinal momentum
distribution ($p_{t}$, $p_{l}$) have been used for identification
of $\pi^{+}$ mesons. It had been assumed, that
$\pi^{-}$ and $\pi^{+}$ mesons hit a given cell of the plane
($p_{t}$, $p_{l}$) with equal probability.
The difference in multiplicity of $\pi^{+}$  and $\pi^{-}$
in each event was required to be no more than 2. After
this procedure the admixture of $\pi^{+}$ is less than 
(5-7)$\%$. The temperature of the identified protons agrees with our
previous result \cite{chkha}, obtained by the spectra subtraction.

In the transverse momentum analysis technique of P.Danielewicz and G.Odyniec
\cite{pawel}
the reaction plane is defined by the transverse vector
$\overrightarrow{Q}$
\begin{eqnarray}
\overrightarrow{Q}=\sum\limits_{i=1}\limits^{n}\omega_{i}
\overrightarrow{p_{{\perp}i}}
\end{eqnarray}
where $i$ is a particle index, $\omega_{i}$ is a weight and $p_{{\perp}i}$
is the transverse momentum of particle $i$.
The reaction plane is the plane containing $\overrightarrow{Q}$
and the beam axis. The weight $\omega_{i}$
is taken as 1 for y$_{i}$$>$ y$_{cm}$ and -1 for y$_{i}$$<$ y$_{cm}$,
where y$_{cm}$ is c.m.s. rapidity and y$_{i}$ is the rapidity of
particle $i$. Autocorrelations are removed
by calculating $\overrightarrow{Q}$ individually for each
particle without including that
particle into the sum (1).
\begin{eqnarray}
\overrightarrow{Q_{j}}=\sum\limits_{i\not=j}\limits^{n}\omega_{i}
\overrightarrow{p_{{\perp}i}}
\end{eqnarray}
The transverse momentum of each
particle  in the estimated reaction plane is calculated as
\begin{eqnarray}
p_{xj}\hspace{0.01cm}^{\prime} = \frac{
\sum\limits_{i\not=j} \omega_{i} \cdot (
\overrightarrow{p_{{\perp}j}} \cdot \overrightarrow{p_{{\perp}i}})}
{\vert\overrightarrow{Q_{j}}\vert}
\end{eqnarray}
As we study an asymmetric pair of colliding nuclei, we chose to bypass the
difficulties associated with the center-of-mass determination and carried
out the analysis in the laboratory frame. We have replaced the
original weight $\omega_{i}$, by the continuous function
$\omega_{i}$= $y_{i}$ - $<y>$ as in \cite{beav}, where  $<y>$ is the
average rapidity, calculated for each event over all the participant protons.

It is known \cite{pawel}, that the estimated reaction plane
differs from the true  one,
due to the finite number of particles in each event.
For the estimation of reaction plane resolution we divided
randomly each event into two equal sub-events (1 and 2) and the
reaction plane vector in both half-events have been evaluated
separately, getting $\overrightarrow{Q_{1}}$ and $\overrightarrow{Q_{2}}$. Then
estimated the azimuth angle difference between the vectors
$\overrightarrow{Q_{1}}$ and $\overrightarrow{Q_{2}}$  $\Delta\varphi_{R}=
\Phi_{1}-\Phi_{2}$. The distributions of this $\Delta\varphi_{R}$
have been plotted for the whole experimental samples.
 The distributions are sharply peaked around $\Delta\varphi_{R}=0$ ( for
C-Cu see Fig. 2).
A uniform distributions have been recovered
for the whole event ($\Phi$) and two sub-events
($\Phi_{1}$, $\Phi_{2}$) reaction planes azimuth angles.  According to
ref. \cite{pawel}, the $\sigma_{r}\equiv<\Delta\varphi_{R}>/2$ is a good 
measure of the reaction plane resolution. We found that $\sigma_{r}=26^{0}$ for C-Ne and
$\sigma_{r}=23^{0}$ for C-Cu, respectively.

First, in Fig. 3 we examine the transverse momentum spectra of protons detected on
the same $(dN^{+}/dp_{t})$ and opposite $(dN^{-}/dp_{t})$ side of the
transverse flow (reaction plane) in central C-Ne (window a) and C-Cu (window b)
interactions, respectively. These protons are not used in the definition
of the reaction plane.
Here only protons emitted in the rapidity intervals
of 1.7$<$ Y $<$ 2.4 for C-Ne
and 1.3$<$ Y $<$ 2.7 for C-Cu, respectively, are considered.
The chosen rapidity ranges are around the projectile rapidity of
$Y_{proj}=2.28$. In these rapidity ranges particles with high transverse
momenta must have suffered very violent collisions and thus originate
most likely from the very hot and dense participant region.
On the other hand, particles with low transverse
momenta are mostly from cold spectators. It is seen that there is a clear excess
of protons emitted to the same side of the directed flow for both C-Ne and C-Cu collisions.
In both collisions the spectra show typical exponential behaviour for
p$_{t} \geq $ 0.2 GeV/c. The spectra for particles in
the same and opposite directions of the transverse flow are approximately
parallel to each other at p$_{t}$ larger than about 0.7 GeV/c. These
findings
are similar to those observed by the E877 collaboration for protons in central Au-Au
collisions at a beam energy of 10.8 GeV/nucleon\cite{e877}.

The above observations can be understood qualitatively within the transversely moving thermal
model of Li et al.\cite{bali3}. Assuming that all or a fraction of
particles in a small rapidity bin around $Y_{proj}$ are in local thermal
equilibrium at a local temperature $T$, and the center of mass of these
particles are moving with a velocity $\beta$ along the
transverse flow direction $+x$ in the reaction plane, then
the transverse momentum spectrum of these particles can be written as
\begin{eqnarray}
\frac{d^{3}N}{p_{t}dp_{t}d \phi dY}=C\cdot \gamma (E- \beta \cdot p_{t}cos(\phi))
\cdot e^{-\gamma (E-\beta p_{t}cos(\phi))/T}
\end{eqnarray}
where $C$ is a normalization constant, $\gamma=1/\sqrt{1-\beta^2}$ and $\phi$
is the azimuthal angle with respect to the reaction plane. The transverse momentum spectra
for particles emitted in the same ($dN^{+}/p_{t}dp_{t}$) and opposite
($dN^{-}/p_{t}dp_{t}$) directions of
the transverse flow are then
\begin{eqnarray}\label{spect}
\frac{dN_{\pm}}{p_{t}dp_{t}}=C_{\pm}e^{- \gamma E/T}(\gamma E \mp T\alpha)
e^{\pm \alpha}
\end{eqnarray}
where $\alpha\equiv\gamma \beta p_{t}/T$. These distributions reduce to simple
exponents
\begin{eqnarray}
\frac{dN_{\pm}}{p_{t}dp_{t}} \propto exp(-p_{t}/T^{\pm}_{eff})
\end{eqnarray}
at high transverse momenta $p_t$. In the above equation, the inverse slopes or effective
temperatures $T^{\pm}_{eff}$ in the semilogarithmic plot of the spectra at high $p_{t}$ are
\begin{eqnarray}
\frac{1}{T^{\pm}_{eff}}= - \lim_{p_{t}\to\infty} \biggl[\frac{d}{dp_{t}}
\cdot ln \Bigl(\frac{dN^{\pm}}{p_{t}dp_{t}} \Bigr) \biggr ]=
\frac{\gamma}{T} \Bigl (cosh (Y) \mp \beta \Bigr).
\end{eqnarray}
In general, the inverse slope $1/T^{\pm}_{eff}$ reflects combined effects of the
temperature $T$ and the transverse flow velocity $\beta$. For the special case
of considering particles at midrapidity, $\beta$ is zero and the effective temperatures
are equal to the local temperature $T$ of the thermal source. Otherwise, one
normally expects $T^{+}_{eff}>T^{-}_{eff}$. For high energy heavy ion collisions
in the region of 1 to 10 GeV/nucleon, $\beta$ is much smaller than $cosh(Y)$
around the projectile rapidity, thus one again expects $T^{+}_{eff}
\approx T^{-}_{eff}$ at high transverse momenta, i.e two approximately parallel spectra
$dN^{+}/p_{t}dp_{t}$ and $dN^{-}/p_{t}dp_{t}$ at high $p_t$.
The distributions $(dN^{+}/dp_{t})$ and  $(dN^{-}/dp_{t})$ have been fitted
by eq.(5). For C-Ne in the rapidity interval of 1.7$<$ Y $<$ 2.4
 from the $(dN^{+}/dp_{t})$ spectra at high $p_{t}$ ( $p_{t} >0.6$ GeV/c) the
temperature $T$  and flow velocity have been obtained:
$T=128 \pm 5$
MeV/c, $\beta=0.016 \pm 0.004$; from the $(dN^{-}/dp_{t})$ spectra at high $p_{t}$ the 
 following values of temperature $T$ and flow velocity have been obtained: $T=130 \pm 4$ MeV/c,
 $\beta=0.019 \pm 0.005$. For C-Cu in the rapidity interval of 1.3$<$ Y $<$ 2.7 respectively,
from the $(dN^{+}/dp_{t})$ spectra at high $p_{t}$ ( $p_{t} >0.6$ GeV/c):
$T=135 \pm 4$ MeV/c, $\beta=0.025 \pm 0.006$;
from the $(dN^{-}/dp_{t})$ spectra:
$T=132 \pm 6$ MeV/c, $\beta=0.023 \pm 0.007$.
The results of fitting are superimposed on spectra in Fig. 3.
One can see, that the values  of temperatures and flow velocities from the
$(dN^{+}/dp_{t})$ and $(dN^{-}/dp_{t})$ spectra coincide both for C-Ne and C-Cu.

To study the strength of differential transverse flow, we compare
in Fig. 4 the ratios $R(p_{t})$ as a function of p$_{t}$ for C-Ne and
C-Cu collisions. It is seen that the ratios increase gradually at low p$_{t}$
and reach a limiting value of about 1.9 and 2.5 in the reaction of C-Ne
and C-Cu, respectively. The values of $R(p_t)$ are greater than one in the
whole transverse momentum range, indicating that protons are emitted preferentially
in the flow direction at all transverse momenta. It is an
unambiguous signature of the sideward collective flow.
Similar results have also been obtained by the E877 collaboration for
protons in Au-Au collisions at 10.8 GeV/nucleon where the ratio
$R(p_{t})$ increases with $p_{t}$ and finally saturates at about 2.

The saturation of $R(p_t)$ at high $p_t$ is also what one expects
within the transversely moving thermal model. From Eq. \ref{spect}
one obtains readily
\begin{eqnarray}
R(p_{t})=\frac{dN^{+}/p_{t}dp_{t}}{dN^{-}/p_{t}dp_{t}}=
\frac{C_{+}}{C_{-}} \cdot \frac{1- \beta \cdot \frac{p_{t}}{E}}
{1+ \beta \cdot \frac{p_{t}}{E}} \cdot e^{2p_{t} \cdot \frac{\gamma\beta}
{T}}.
\end{eqnarray}
This ratio normally increases with $p_{t}$ for any given values of $T$ and $\beta$,
but it becomes almost a constant at high $p_t$ for very small $\beta/T$ ratios.
In relativistic heavy-ion collisions, particles are emitted continuously at different
freeze-out temperatures during the whole reaction process. In particular,
particles with high $p_{t}$ are mostly emitted in the early stage of the reaction
from the most violent regions where the local temperatures are high. Since the
ratio $R(p_{t})$ varies very slowly with $p_t$ for low $\beta/T$ ratios,
one thus expects an approximately constant value of $R(p_{t})$ at high $p_t$.
At low transverse momenta, however, $R(p_{t})$ is affected mostly
by particles from the cold spectators, and the ratio $R(p_t)$ approaches
one as $p_{t}$ goes to zero.

Now we turn to the analysis of differential transverse flow for negative pions
in the C-Ne and C-Cu reactions. Shown in Fig. 5 are the $R(p_{t})$ ratios
for negative pions in the rapidity range of $1.0 < Y< 2.5$. The solid lines are
the linear fits to the experimental data to guide the eye.
In the C-Ne collisions the ratio $R(p_t)$ is greater than one in the whole range of
transverse momentum, indicating pions are flowing on average to the same direction as protons.
Moreover, the strength of the pion differential transverse flow increases
almost linearly with $p_{t}$ and reaches about 1.2 at $p_{t}$=0.9 GeV/c.
On the contrary, it is seen that an anti-flow is observed for $\pi^{-}$ mesons
in the reaction of C-Cu. Thus, in C-Ne interactions the pions are preferentially
emitted in the direction of the proton transverse flow, while in C-Cu reactions pions are
preferentially emitted away from the direction of the proton transverse flow. It is necessary
to mention here that a similar target dependence of the pion preferential emission in asymmetric
nucleus-nucleus collisions was first observed by the DIOGENE
collaboration\cite{dio}. Our results are consistent with theirs.
This target dependence of the apparent pion transverse flow was explained quantitatively by
nuclear shadowing effects of the heavier target within nuclear transport models \cite{bali4,bass}.
Our results on the pion differential transverse flow and its dependence on the target are thus
also understandable.

\section{SUMMARY}

In summary, the differential transverse flow of protons and pions in
central C-Ne and C-Cu collisions at a beam energy of 3.7 Gev/nucleon were measured
at the SKM-200-GIBS setup of JINR. The strength of proton differential
transverse flow is found to first increase gradually and then saturate with
the increasing transverse momentum.
>From the transverse momentum distributions of protons emitted in the
reaction plane to the same  and opposite  side of the transverse
flow the temperatures and flow velocity $\beta$ have been extracted
in C-Ne  and C-Cu collisions.
The observations are in qualitative
agreement with predictions of a transversely moving thermal model.
In the whole range of transverse momentum studied, pions are found to be
preferentially emitted in the same direction of the
proton transverse flow in the reaction of C-Ne, while an anti-flow of pions is found in the
reaction of C-Cu due to stronger shadowing effects of the heavier target.
The differential flow data for both protons and pions provide a more
stringent testing ground for relativistic heavy-ion reaction theories.
Detailed comparisons with predictions of relativistic transport models
on the differential flow will be useful to extract more reliable information
about the nuclear EOS. Such an endeavor is underway and results will be published
elsewhere.

\section{ACKNOWLEDGEMENT}

We would like to thank M. Anikina, A. Golokhvastov, S. Khorozov and  J. Lukstins
for fruitful collaboration during the obtaining of the data.
The work of B.A. Li was supported in part by the U.S. National Science
Foundation under Grant No. PHY-0088934 and the Arkansas Science and Technology
Authority under Grant No. 00-B-14.

\newpage
\section*{Figure Captions}
\begin{description}
\item {Fig. 1.} SKM-200-GIBS experimental set-up. The trigger and the
trigger distances are not to scale.

\item {Fig. 2.} Reaction plane resolution for C-Cu collisions. The curve
is the result of data approximation by the polynom of 4-th order to guide
the eye.

\item {Fig. 3.} The transverse momentum distributions of protons emitted in the
reaction plane to the same ($\circ$) and opposite ($\bigtriangleup$) side of the transverse
flow in a) C-Ne  and b) C-Cu collisions. The lines are results of fitting by
eq.(5) at high $p_t$.

\item {Fig. 4.} Ratios of the yield of protons emitted
to the same and opposite side of the transverse flow as a function of $p_{t}$ in
C-Ne ($\circ$) and  C-Cu ($\bigtriangleup$) collisions. The curves are logarithmic
fits to guide the eye.

\item {Fig. 5.} Ratios of the yield of $\pi^{-}$ mesons emitted
to the same and opposite side of the transverse
flow as a function of $p_{t}$ in  C-Ne ($\circ$) and  C-Cu ($\bigtriangleup$)
collisions. The lines are linear fits to the data to guide the eye.

\end{description}
\newpage
\begin{figure}[htp]
\vspace{-1.09 cm}
\begin{center}
\epsfig{file=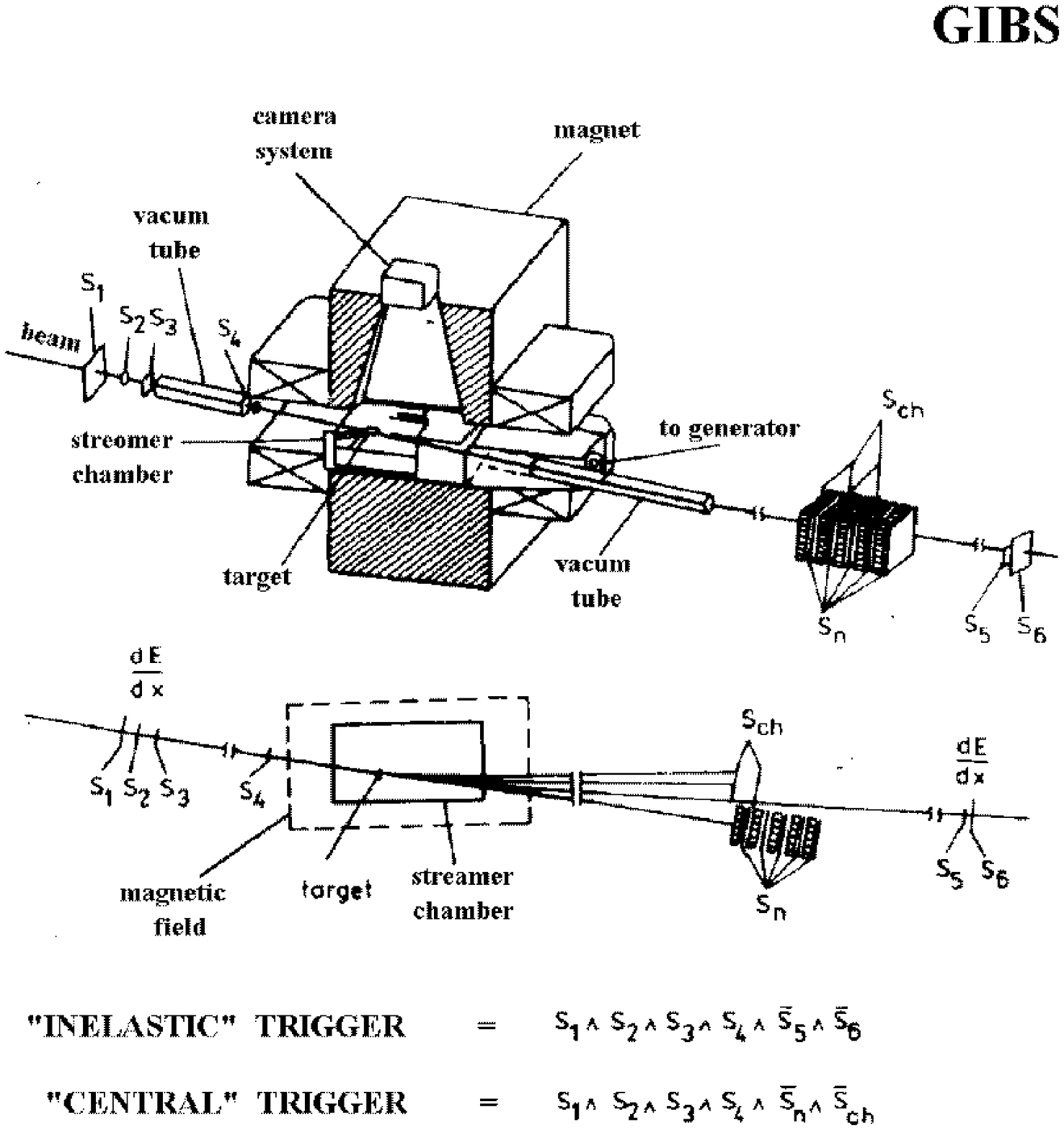,bbllx=0pt,bblly=10pt,bburx=594pt,bbury=720pt,
width=18cm,angle=0}
\vspace{-4.5cm}
\caption
{ Experimental set-up. The trigger and trigger distances are not scale.}
\end{center}
\end{figure}
\newpage
\begin{figure}[htp]
\vspace{-1.09 cm}
\begin{center}
\epsfig{file=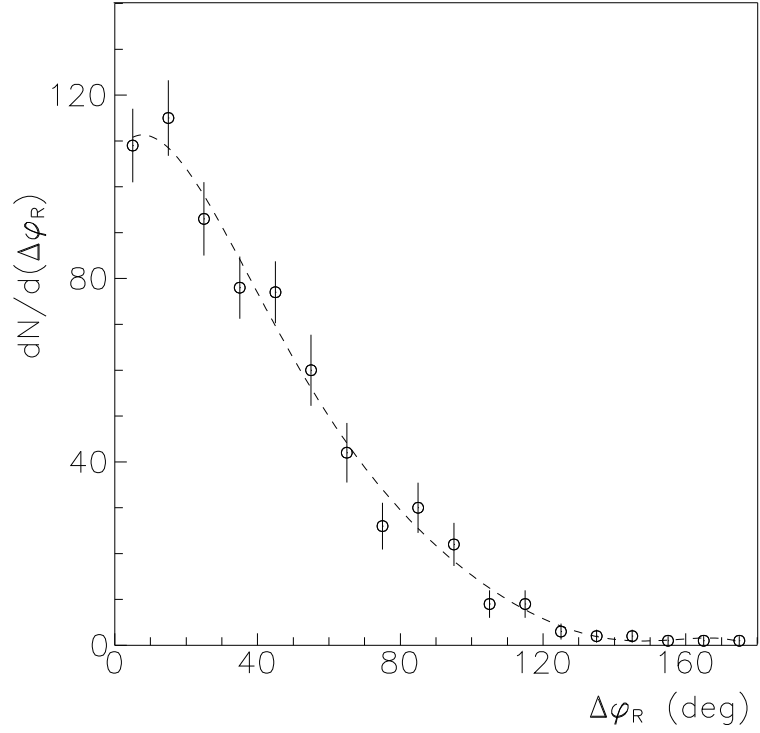,bbllx=0pt,bblly=10pt,bburx=594pt,bbury=720pt,
width=18cm,angle=0}
\vspace{-7.cm}
\hspace{-4.cm}
\caption
{ Reaction plane resolution for C-Cu collisions. The curve
is the result of data approximation by the polynom of 4-th order to guide
the eye.}
\end{center}
\end{figure}
\newpage
\begin{figure}[htp]
\vspace{-1.09 cm}
\begin{center}
\epsfig{file=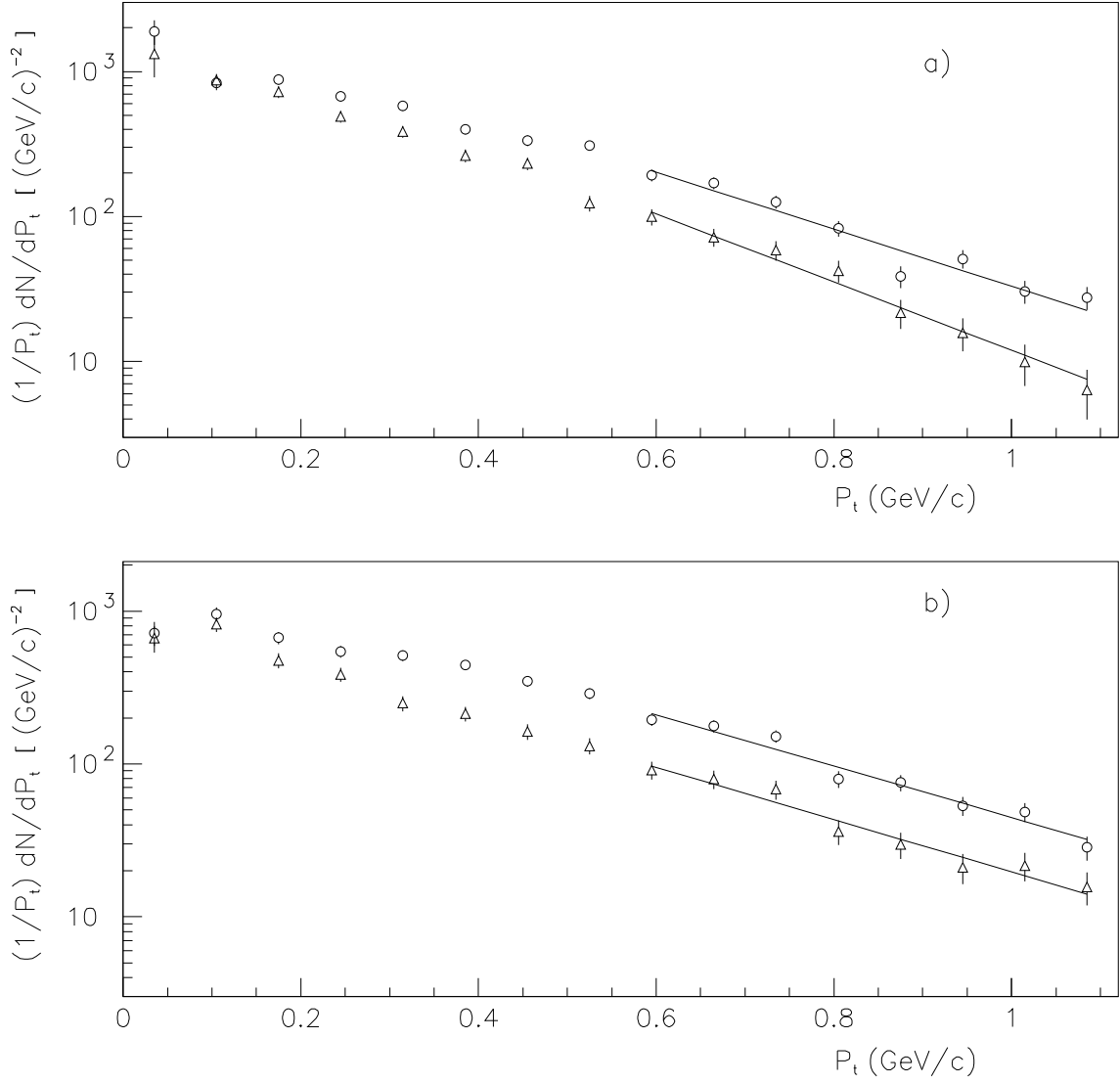,bbllx=0pt,bblly=10pt,bburx=594pt,bbury=720pt,
width=18cm,angle=0}
\vspace{-5.cm}
\caption
{ The transverse momentum distributions of protons emitted in the
reaction plane to the same ($\circ$) and opposite ($\bigtriangleup$) side of the transverse
flow in a) C-Ne  and b) C-Cu collisions.
 The lines are results of fitting by
eq.(6) at high $p_t$.}
\end{center}
\end{figure}
\newpage
\begin{figure}
\begin{center}
\epsfig{file=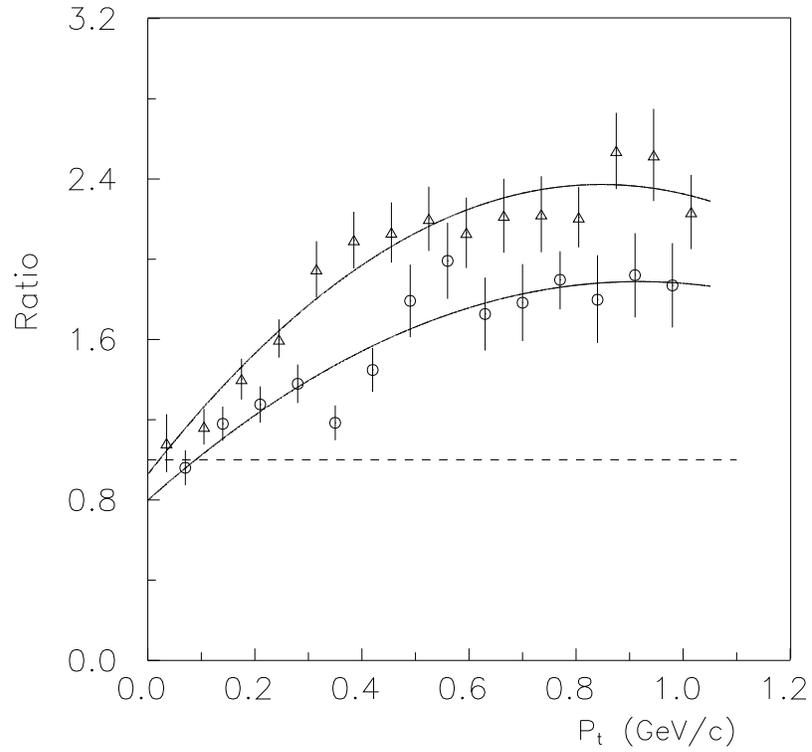,bbllx=0pt,bblly=0pt,bburx=594pt,bbury=720pt,
width=18cm,angle=0}
\vspace{-8.cm}
\hspace{0.cm}
\caption
{ Ratios of the yield of protons emitted
to the same and opposite side of the transverse
flow as a function of $p_{t}$ in  C-Ne ($\circ$) and  C-Cu ($\bigtriangleup$)
collisions. The curves are logarithmic fits to the data to guide the eye.}
\end{center}
\end{figure}
\newpage
\begin{figure}
\begin{center}
\epsfig{file=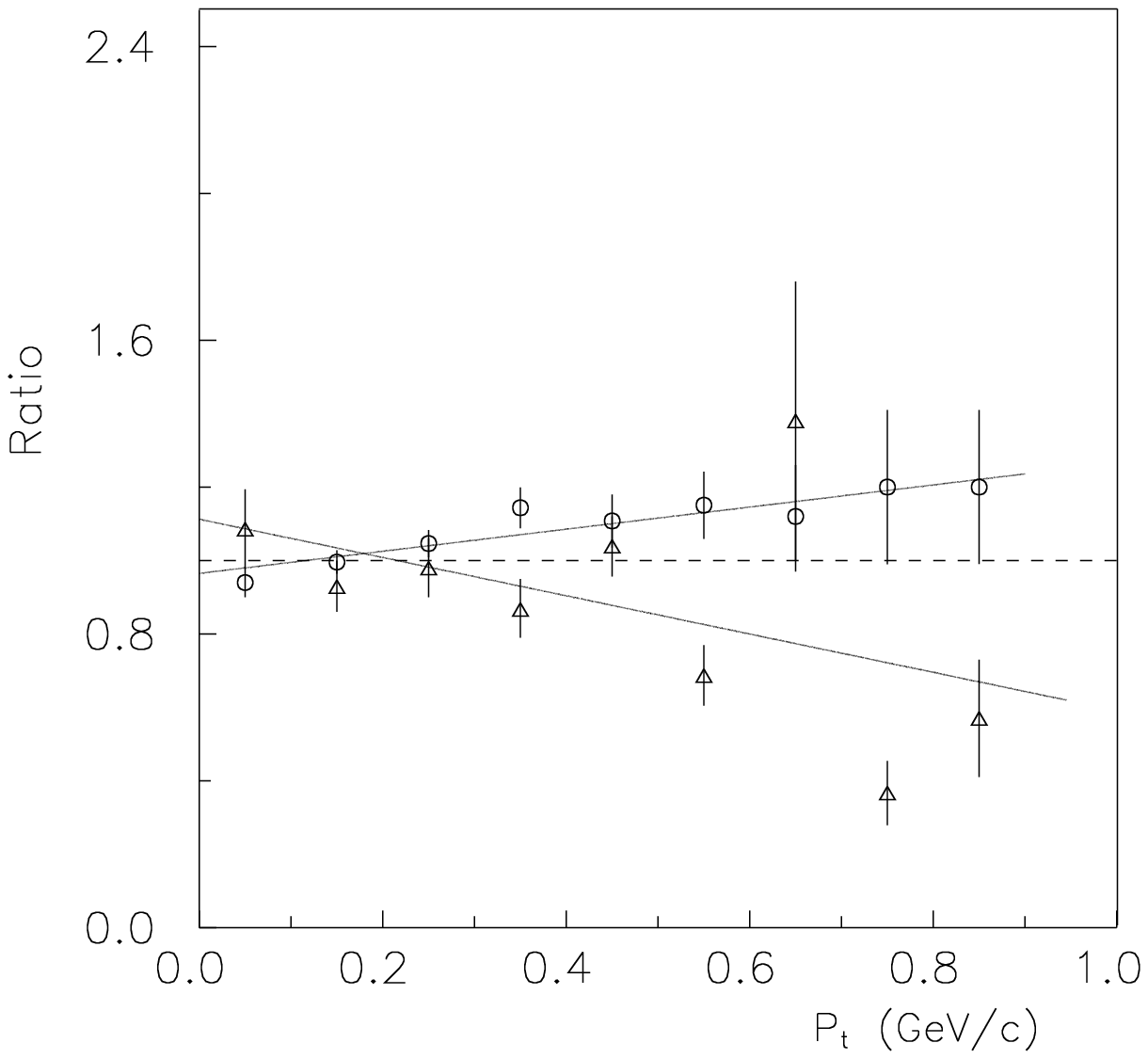,bbllx=0pt,bblly=0pt,bburx=594pt,bbury=720pt,
width=18cm,angle=0}
\vspace{-8.cm}
\hspace{0.cm}
\caption
{ Ratios of the yield of $\pi^{-}$ mesons emitted
to the same and opposite side of the transverse
flow as a function of $p_{t}$ in  C-Ne ($\circ$)
and  C-Cu ($\bigtriangleup$)
collisions. The lines are linear fits to the data to guide the eye.}
\end{center}
\end{figure}

\newpage
\begin{thebibliography}{99}
\bibitem{gary}G.D. Westfall, Nucl. Phys. {\bf A630}, 27c (1998);
{\it ibid}, {\bf A681}, 343c (2001).

\bibitem{hhg}H.H. Gutbrod, A.M. Poskanzer and H.G. Ritter,
Rep. Prog. Phys. {\bf 52}, 1267 (1989).

\bibitem{reis}W. Reisdorf and H.G. Ritter,
Ann. Rev. Nucl. Part. Sci. {\bf 47}, 663 (1997);
N. Hermann, J.P. Wessels and T. Wienold, {\it ibid}, {\bf 49}, 581 (1999).

\bibitem{ogi}C.Ogilvie et.al., Nucl.Phys., {\bf A638}, 57 (1998).

\bibitem{rai}G. Rai, Nucl. Phys. {\bf A681}, 181 (2001).

\bibitem{rhic}H.Appelshauser et. al.(NA49 collaboration),
Phys. Rev. Lett., {\bf 80}, 4136 (1998);
M.M.Aggarwal et. al., (WA98 collaboration), Phys. Lett. {\bf B469}, 30 (1999);
K.Ackermann et. al. (STAR collaboration), Phys. Rev. Lett., {\bf 86}, 402 (2001);
K.Adcox et. al.(PHENIX Collaboration), Phys. Rev. Lett. {\bf 87}, 052301 (2001).

\bibitem{eos}C. Pinkenburg et al. (E895 collaboration),
Phys. Rev. Lett. {\bf 83}, 1295 (1999);  H.Liu et.al.(E895 collaboration),
{\it ibid}, {\bf 84}, 548 (2000).

\bibitem{e917}B. Holzman et al (E917 Collaboration),
Nucl-ex/0103015, in QM01 Proc., Nucl., Phys. A (2001) in press.

\bibitem{dubna1}L. Chkhaidze, T.Djobava et al., Phys. Lett., {\bf B411}, 26 (1997).
\bibitem{dubna2}L. Chkhaidze, T.Djobava et al., Phys. Lett., {\bf B479}, 21 (2000).
\bibitem{dubna3}Lj. Simic and J. Milosevic, J. Phys., {\bf G27}, 183(2001).
\bibitem{dubna4}B. Bannik et al., J. Phys., G14, 949 (1988).
\bibitem{dubna5}M. Adamovich et al., Eur. Phys. J., {\bf A6}, 427 (1999).

\bibitem{exp1}M. Anikina et al., JINR Dubna Report E1-84-785 (1984).

\bibitem{exp2}M. Anikina et al., Phys. Rev. C{\bf 33}, 895(1986).

\bibitem{gud}K. Gudima, V.Toneev, Nucl. Phys.  A{\bf 400}, 895(1983).

\bibitem{ben}O. Benary, R.Price G. Alexander, UCRL -20000 NN report, 1970.

\bibitem{anik}M. Anikina et al., Z. Phys. C{\bf 25}, 1(1984).

\bibitem{pawel}P. Danielewicz and G. Odyniec, Phys. Lett., {\bf B157}, 146 (1985).

\bibitem{ell1}J.-Y. Ollitrault, Phys. Rev. D{\bf 46}, 229 (1992).

\bibitem{ell2} S. Voloshin and Y. Zhang, Z. Phys., C{\bf 70}, 665 (1996);
  A.M. Poskanzer and S.A. Voloshin, Phys. Rev., C{\bf 58}, 1671 (1998).

\bibitem{bali1}B.A. Li and A.T. Sustich, Phys. Rev. Lett. {\bf 82}, 5004 (1999);
B.A. Li, {\it ibid}, {\bf 85}, 4224 (2000).

\bibitem{bali2}B.A. Li, A.T. Sustich and B. Zhang, nucl-th/0108047,
Phys. Rev. C{\bf 64}, 054604 (2001).

\bibitem{gyu}M. Gyulassy, I. Vitev, X.-N. Wang, Phys. Rev. Lett. {\bf 86}, 2537 (2001).

\bibitem{fopi}A. Andronic et al. (FOPI Collaboration), 
Phys. Rev. C{\bf 64}, 041604 (2001).

\bibitem{e877}J. Barrette et al. (E877 collaboration),
Nucl. Phys. {\bf A590}, 259c (1995).

\bibitem{bali3}B.A. Li, C.M. Ko and G.Q. Li, Phys. Rev. C{\bf 55}, 844(1996).

\bibitem{vio}S.A. Voloshin, Phys. Rev. C{\bf 55}, R1630 (1997).

\bibitem{chkha}L. Chkhaidze al., Z. Phys. C{\bf 54}, 179(1992).

\bibitem{beav}O. Beavis, Phys. Rev. C{\bf 45}, 299 (1992).

\bibitem{dio}J. Gosset et al. (DIOGENE Collaboration),
Phys. Rev. Lett., {\bf 62}, 1251 (1989);
M. Demoulins et al. (DIOGENE Collaboration), Phys. Lett., {\bf B241}, 476 (1990).

\bibitem{bali4}B.A. Li, W. Bauer and G.F. Bertsch,
Phys. Rev., C{\bf 44}, 2095 (1991); B.A. Li, Nucl. Phys. {\bf A570}, 797 (1994).

\bibitem{bass}S.A. Bass, R. Mattiello, H. St\"ocker, W. Greiner and
C.Hartnack, Phys. Lett., {\bf B302}, 381 (1994).
\end{thebibliography}
\end{document}